# A Natural Language Processing Pipeline for Detecting Informal Data References in Academic Literature

**Lafia, Sara**     ICPSR, University of Michigan, Ann Arbor, MI, USA | slafia@umich.edu

**Fan, Lizhou**     School of Information, University of Michigan, Ann Arbor, MI, USA | lizhouf@umich.edu

**Hemphill, Libby**     School of Information, University of Michigan, Ann Arbor, MI, USA| libbyh@umich.edu

## ABSTRACT

Discovering authoritative links between publications and the datasets that they use can be a labor-intensive process. We introduce a natural language processing pipeline that retrieves and reviews publications for informal references to research datasets, which complements the work of data librarians. We first describe the components of the pipeline and then apply it to expand an authoritative bibliography linking thousands of social science studies to the data-related publications in which they are used. The pipeline increases recall for literature to review for inclusion in data-related collections of publications and makes it possible to detect informal data references at scale. We contribute (1) a novel Named Entity Recognition (NER) model that reliably detects informal data references and (2) a dataset connecting items from social science literature with datasets they reference. Together, these contributions enable future work on data reference, data citation networks, and data reuse.

## INTRODUCTION

Recent work investigating the relationship between properties of research datasets, curation decisions, and reuse has found that intensively curated data tend to be used more often (Hemphill et al., 2021). However, there is much more to learn about the context of research data reuse, such as the academic disciplines that gravitate toward particular datasets and the variety of applications in which data are used. Analyzing data citations can shed light on data use. However, discovering links between publications and the datasets they use is non-trivial as many authors informally describe data that they have used rather than formally citing the data (Mooney, 2011).

As a consequence, retrieving comprehensive references to data can be a labor-intensive process. At ICPSR, a large social science data archive, staff curate a bibliography of data-related literature linking 11,000 social science studies to 100,000 data-related publications that use them. Given limited resources at data archives like ICPSR, priority is often given to monitoring the use of high profile datasets, such as well-established and long-running studies, rather than lower-profile datasets in the "long tail" (Heidorn, 2008). As a result, the citation counts constructed from manual efforts tend to be conservative estimates, which undercount the degree of actual data usage.

Our work contributes to the study of research data citation and reuse in two main ways. First, we developed a natural language processing pipeline to detect informal data references in academic literature. By automating the detection of data references, we complement data librarians' manual work to identify connections between data and publications. To accomplish detection, we trained a custom Named Entity Recognition (NER) model on a dataset of several thousand manually verified data citation sentences (Lafia et al., 2021). The model detects informal dataset references in academic text with relatively high recall. Second, we applied the trained pipeline in a human-in-the-loop paradigm to expand the ICPSR Bibliography of Data-Related Literature (Fan et al., 2022). This expanded dataset enables research on data reference, data citation networks, and data reuse. Together, the automation and expansion steps in the pipeline increased ICPSR's capacity to identify connections between datasets and research publications across a vast body of literature, complementing ongoing labor by staff to track research data citations.

### Data citation practices

Data citation practices in the social sciences are evolving. The assignment of machine-readable, persistent identifiers (PIDs) to research data is an emerging standard that supports the disambiguation and automated tracking of data citations (Cousijn et al., 2019). Data citation gives credit to research data, enabling the study of research data indicators and impact (Buneman et al., 2022). However, there is still a great deal of literature in which data references are vague, incomplete, and indirect (Moss & Lyle, 2018). This is especially true for the retrospective analysis of data use prior to the advent of persistent identifiers for data. Informal references are time consuming to address and require human interpretation to infer which datasets the authors have used (Yarkoni et al., 2021).

Data citations differ from other types of citations in several ways. Unlike references to literature, which are typically standardized and included in a works cited section, statements indicating data use tend to occur throughout the body of text, tables, footnotes, and acknowledgements sections (Mayo et al., 2016). A study of the Dryad digital repository found that while the share of publications citing persistent identifiers for data had grown from 69% to 83% between 2011 and 2014, the share of publications that included data identifiers in their works cited section remained low, under 10% (He & Han, 2017). In a study of biomedical research articles, most data use indicators were informal and did not reference datasets' persistent identifiers; these "mentions" of data were found mainly in the materials, acknowledgements, and supplemental materials sections of published articles (Park et al., 2018).

## Analyzing data citations

Capturing and analyzing data citations is essential for understanding the context of data reuse and informing decisions about data governance (Hellerstein et al., 2017). However, data citation practices are inconsistent. Detecting informal data references in literature could provide a more complete view of data reuse. Data providers are leveraging machine learning to address the disconnect between research data and publications. Named entity recognition (NER) has been proposed to find references to research data but has achieved low recall at the level of individual dataset mentions and documents (Boland et al., 2012). Recent work in Artificial Intelligence has achieved substantially higher recall by applying deep-learning, transformer models, like BERT, to NER tasks to identify references to machine learning datasets (Heddes et al., 2021). Similar efforts are also being developed to detect mentions of software using combinations of NER and rule-based extraction methods to identify the "invisible" contributions that software makes to academic research (Du et al., 2021).

Many information retrieval tasks, such as document summarization and semantic analysis, rely on the analysis of citation-bearing sentences, or "citances" (Nakov et al., 2004). Scientific citations serve many functions such as providing background, motivation, use cases, extensions, comparisons, or proposals of future work (Jurgens et al., 2018). In particular, data citations indicate the disciplinary reach of scientific datasets, making it possible to study the impact of data products on academic disciplines (Chao, 2011).

## DATA DETECTION PIPELINE

We developed a computational pipeline to detect informal data references in academic literature (**Figure 1**). This pipeline includes three steps: retrieve, parse, and predict. The pipeline accepts a catalog of datasets in the ICPSR Bibliography and automatically generates a set of sentences in which informal data references are tagged.

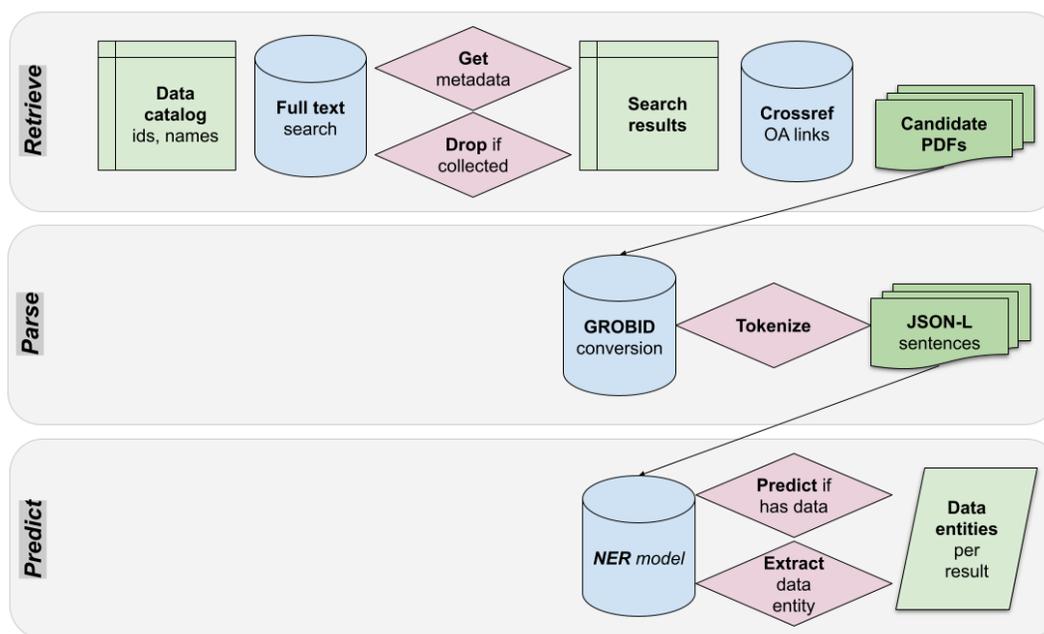

**Figure 1. Overview of inputs, outputs, and services used in the data detection pipeline**

The pipeline starts from a data catalog and searches for dataset identifiers (digital object identifiers, archival study numbers) and names (canonical, variants) in a full text index of publications. We use the Dimensions database to search against an index of more than 69 million full text publications (Hook et al., 2018). Access to the Dimensions platform is provided by Digital Science through subscription-only data sources under a license agreement through the University of Michigan. From the Dimensions database, the pipeline retrieves metadata for candidate full text publications, which contain one or more dataset identifiers or names and which have not yet been collected in the ICPSR Bibliography. Metadata for candidate documents is then passed to the Crossref API, which is used to request open access full text PDFs through content negotiation (Lammey, 2015).

Next, the pipeline parses full text PDFs into structured text documents using a service called GROBID (Lopez, 2009). This conversion method retains headers, section titles, tables, figures, and footnotes where data references are likely to be found. The documents are then tokenized with pySBD, a pragmatic sentence boundary disambiguation library for scientific texts (Sadvilkar & Neumann, 2020).

Finally, the pipeline applies a custom Named Entity Recognition (NER) model to each sentence in a document to detect dataset entities. We describe the creation of training data for our NER model and evaluate the model's performance in the next section. The pipeline outputs a set of data entities and the sentences that contain them as predicted data references for each document. These predictions support bibliometric ingest workflows by summarizing informal data references across many documents and help prioritize documents to review.

## Creating training data

To detect informal data references in academic literature, we first developed a training dataset for an NER model. We used the following definition of a *dataset* from the ICPSR Glossary of Social Science Terms: "collection of data records… numeric files originating from social research methodologies or administrative records, from which statistics are produced". As an example, this bolded mention indicates a dataset: "We also investigate individual-level black-white thermometer scores from waves of the **American National Election Survey (ANES)** from 1984 until 1998." Individuals labeling data for the training set used following guidelines:

- Only label *dataset* proper names ("Housing and Urban Development (HUD): Family Unification Program Youth Data") rather than general references to data products ("a 1994 HUD survey….")
- If the *dataset* span is contiguous, keep it intact ("Early Childhood Longitudinal Study (ECLS)")
- If a year is given as part of the proper *dataset* name, include it in the label ("2007 ARMS Phase III")
- Omit trailing words that are not part of the proper name of the *dataset* ("ECLS-K" study, survey)
- Label all references to a *dataset* even if mentioned more than once ("2013 LEMAS; LEMAS")

We selected training sentences from two sources of academic research publications: 1) the ICPSR Bibliography of Data-Related Literature (Moss et al., 2015); and 2) the Semantic Scholar Open Research Corpus (S2ORC) (Lo et al., 2020). More than half of the items in the ICPSR Bibliography are social science journal articles, followed by theses, reports, book chapters, and other types of materials that use ICPSR data. The S2ORC corpus contains over 8 million full text academic articles from medicine, biology, and other STEM fields, complementing social science literature from the ICPSR Bibliography.

To select a set of training sentences from these documents, we applied morphological patterns inspired by existing rule-based methods (Heddes et al., 2021) and included keywords indicating data use (Park et al., 2018). We created a three-level candidate extraction method for sentences, which is summarized in **Table 1**. These patterns filter the full text and limit review to sentences with at least one matching pattern. In addition, this method specifies levels of possibilities for the annotator to review in the annotation interface, which reduces the amount of processing time by specifying the spans for the dataset reference candidates. We designate sentences that contain an indicator keywords as highly likely to contain a dataset reference, while those that contain an acronym are moderately likely, and those that contain a sequence of upper cased words have a lower likelihood. Through this process, the annotator can simply make judgment about whether an entity span is a true dataset reference, and focus on candidates with higher possibilities. In total, we extracted and labeled 4,505 training sentences: 3,004 sentences came from the ICPSR Bibliography and 1,503 came from S2ORC.

| Level | Rule | Pattern |
|-------|------|---------|
| High | keywords and their variations matched using regular expressions | `'(?:train\|test\|validation\|testing\|trainings?)\s*(?:` `set)', 'data', 'data\s*(?:set\|base)s?',` `'corp(us\|ora)', 'tree\s*bank', 'collections?',` `'benchmarks?', 'surveys?', 'samples?',` `'stud(y\|ies)', 'reports?', 'census(es)?'` |
| Medium | regular expression to return acronyms of names of a dataset | `\b[A-Z]{3,}s?\b` |
| Low | regular expression to return the full names of a dataset | `([A-Z][a-z]+\s){2,}[A-Z][a-z]+` |

**Table 1. Extraction patterns for creating NER training data**

The training process was iterative and allowed us to update our NER model as we labeled batches of sentences (**Figure 2**). The dashed components in the diagram – labeling sets of sentences and training the model – were updated in each iteration following model evaluation. The sentences, along with the span of the labeled *dataset* references, were provided as inputs to assist subsequent annotation steps.

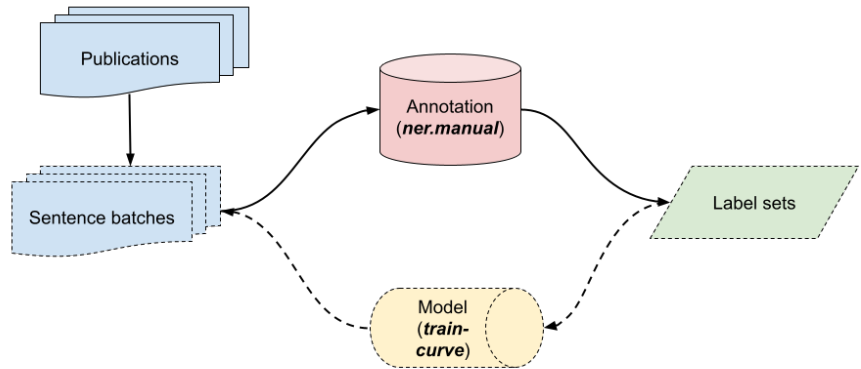

**Figure 2. Overview of workflow for developing NER training data and model**

We used Prodigy, a tool for machine learning and teaching (Montani & Honnibal, 2018) to label *dataset* entities in the extracted sentences. **Figure 3** shows an example of a sentence referring to a dataset, the *National Intimate Partner and Sexual Violence Survey*, highlighted in yellow. The metadata included below the sentence show which morphological patterns the sentence matched. The keywords "Studies", "data", "samples", and "Survey" are listed in **Table 1** as highly likely matching keywords. The acronym "IPV" is a medium-level possibility, as it has consecutive capital letters and can be an acronym for a dataset. In this sentence however, "IPV" means "Intimate Partner Violence", which is a proper noun phrase of public health. The dataset reference candidate with low-level possibility, "National Intimate Partner" and "Sexual Violence Survey" are both parts of the name of the true dataset reference, which help the annotator to locate where to annotate. As this example shows, the three-level candidate extraction method both automatically extracts data reference candidate sentences and provides context for the annotator, which accelerates the creation of the high-quality training data of the NER model. We describe the NER model training and performance in the following sections.

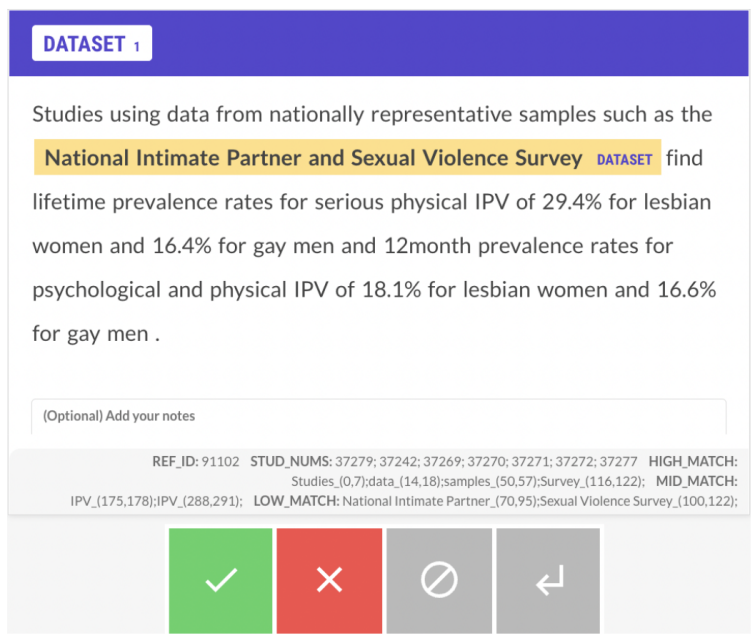

**Figure 3. An annotated training data example for the NER model in the Prodigy interface: a candidate sentence and the extracted dataset reference candidates, with the true dataset reference highlighted in yellow**

### Training a NER model

We split the labeled data into 80% training (3,604 sentences) and 20% evaluation (901 sentences) sets. We labeled additional data and retrained our model until we achieved a target recall score of 80. Recall is the most important measure for our task as we seek to maximize our ability to retrieve all candidate dataset references; we are less concerned with prediction precision because of human-in-the-loop feedback. We used the Python *spaCy* library (Honnibal et al., 2020) to train our NER pipeline with a graphics processing unit using a transformer-based architecture, which relies on the context of tokens to identify entities. Our final model achieved a score of 80.0 for recall, 91.95 for precision, and 85.56 for F1.

### APPLYING THE PIPELINE

ICPSR, a large social science data archive, manually curates a bibliography of data-related literature that links over 11,661 social science studies to 105,797 data-related publications in which they were used. Because the process of finding data citations and verifying data use in literature is labor intensive, the coverage of the ICPSR Bibliography is uneven. For example, the bibliography contains literature published from the inception of ICPSR in 1962 up to the present; however, most items in the bibliography were published after 2000 and reflect more recent scholarship, which tends to be readily discoverable and digitally accessible, rather than older, analog materials (**Figure 4**).

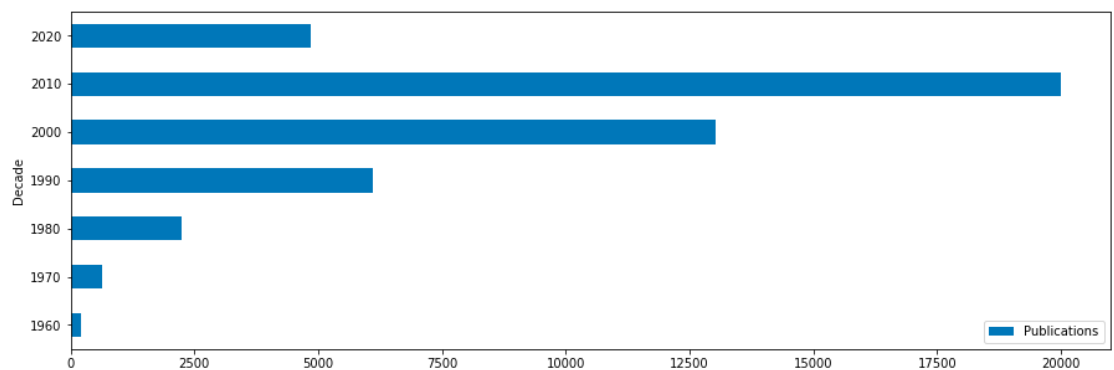

**Figure 4. Publications in the current ICPSR Bibliography grouped by decade**

Data at ICPSR is organized into topical archives, which receive substantially different levels of citation (**Figure 5**). For example, the "demographics" topical archive (Data Sharing for Demographic Research - DSDR) has the highest study-citation ratio of all topical archives (an average of 246 citations/study). It stewards 61 studies, which have collectively received 15,029 total citations. By comparison, the "child and family" topical archive (Child and Family Data Archive - CFDA) also stewards 61 studies, which have received 2,069 total citations (an average of 33.9 citations/study). Citation counts are based on the ICPSR Bibliography. All materials in the bibliography are reviewed by staff to verify that they demonstrate substantial reuse of ICPSR data. Some topical data archives – such as the "demographics", "criminal justice", and "aging" archives – support bibliography staff time, so more materials citing these data tend to be collected. As a result, citations for items in smaller, topical archives that do not pay for bibliography services may be undercounted. We applied our data detection pipeline to expand the current bibliography and balance its coverage by retrieving and reviewing candidate data citations for all ICPSR datasets regardless of their funding source or topic.

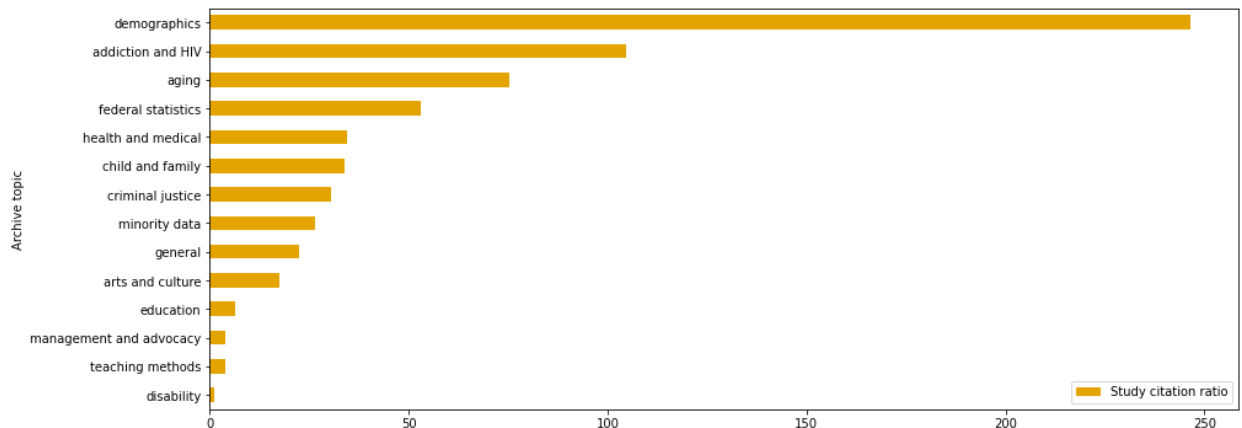

**Figure 5. Average study citations for ICPSR datasets grouped by their topical archive**

### Retrieving data-related publications

We searched for publications referencing ICPSR datasets, which were not part of the current ICPSR Bibliography. At ICPSR, each study consists of one or more data files and metadata; data citations are issued at the study level. We searched all publication full text in the Dimensions database for exact references to all 11,656 ICPSR study names, study DOIs, and study numbers. Our searches excluded studies that were self-deposited by authors in openICPSR. We included studies with public and restricted-use data containing sensitive information. We also included union catalog records, as ICPSR Bibliography staff collect these citations, and included deaccessioned studies, as the associated publications found when the study was active are retained in the bibliography.

We issued the following queries and retrieved results for each type of search query as reported in **Table 2**. ICPSR studies were first assigned DOIs beginning in 2008 so we did not expect to find literature referring to ICPSR DOIs published prior to 2008. Most new results included the exact names of ICPSR studies in their full text (69%), suggesting that many authors refer to data by name but do not formally cite it. Fewer results matched ICPSR study DOIs (17%) or study numbers (14%). We retrieved more instances of authors referring to ICPSR studies by study number, suggesting that fewer authors formally cite ICPSR data. This finding is consistent with prior studies of data citation practices in the social sciences (Moss et al., 2015). We then de-duplicated the search results by comparing search result identifiers to items in ICPSR's Bibliography as of February 2022. After deduplicating the hits against the current bibliography, we retrieved 9,191 new candidate publications that were not part of the current ICPSR Bibliography. If a publication included a dataset study number and a dataset DOI for example, these references were counted separately. We then used the publicly available Crossref API to check the license information and request links to available full text. In all, we retrieved 1,285/9,191 (14%) of the search result candidates as full text PDFs for evaluation. By automating search and review, we were able to identify a wider range of candidate publications for an entire data catalog, ultimately increasing the coverage of the ICPSR Bibliography.

| Search method | Example query string | Total publications | Percent |
|---|---|---|---|
| Study Name | "American Citizen Participation Study 1990" | 6,318 | 69% |
| Study DOI | "10.3886/ICPSR06635" | 1,614 | 17% |
| Study Number | "ICPSR 6635" | 1,259 | 14% |

**Table 2. Search results for publications referencing ICPSR study data**

## Close reading: Using NER to detect dataset references in an academic publication

We use the concepts of "close" and "distant" reading from the digital humanities to illustrate how our model facilitates the detection of informal data references in academic publications (Moretti, 2000). First, we applied the model to a single publication; close reading allows us to assess the quality of the model's predictions by inspecting the context of each extracted data reference. We then applied the model to the large corpus of candidate articles we retrieved to study general trends in data references and to perform distant reading of data references at scale.

We began by coding a research article, The Political Legacy of American Slavery (Acharya et al., 2016), which is included in the ICPSR Bibliography and has been verified to use the following ICPSR datasets: Three Generations Combined, 1965-1997; Youth-Parent Socialization Panel Study, 1965-1997: Four Waves Combined; Historical, Demographic, Economic, and Social Data: The United States, 1790-2002; and the ANES Time Series Study (1984, 1986, 1988, 1990, 1992, 1994, 1996, 1998). We identified 31 sentences in the article that referenced seven distinct datasets or data series. We then applied the NER model to extract data references and compared the model's predictions to the manually verified data references. The model identified 21/31 (68%) of all sentences coded as having data references; however, the NER model successfully identified all seven unique datasets and data series referenced in the paper, demonstrating high recall.

The model also identified several datasets that are not part of ICPSR's data collection: IPUMS data; and the 2010 CCES (Cooperative Congressional Election Study). **Figure 6** shows examples of extracted sentences with predicted data references highlighted in gray. The extracted sentences are a mixture of formal data citations and informal data references or mentions. Formal data citations give credit to the data that the authors analyzed in the references section and cite the dataset using the dataset's lead principal investigator and year. Informal dataset references describe the data used, often by name, without referring to an unambiguous, persistent identifier. While the NER model is designed primarily to capture informal data references, it detects formal data citations as well.

We analyze three county-level outcome measures, which come from the [ Cooperative Congressional Election Study (CCES **DATASET** ) ], a large survey of American adults (Ansolabehere, 2010) .

We pool [ CCES **DATASET** ] data from the 2006, 2008, 2009, 2010 , and 2011 surveys to create a combined data set of over 157,000 respondents.

In addition, we also investigate individual-level black-white thermometer scores from waves of the [ American National Election Survey (ANES **DATASET** ) ] from 1984 until 1998, a time period where the ANES both used a consistent sampling frame and included county-level identifiers for respondents.

**Figure 6. Examples of publication sentences and their predicted dataset references, highlighted in gray**

## Distant reading: Using NER to extract data references from a large corpus of publications

We applied the NER model to the available full text of all search result PDFs (n=1,285) mentioning ICPSR datasets. After tokenizing the publications and applying the NER model, we found that 1.9% (7,486/402,637) of sentences in the sample contained a predicted dataset entity. However, 84% (1,074/1,285) of the publications contained at least one predicted dataset reference. We used a probabilistic record linkage Python library, *fuzzymatcher*, to compare the extracted *dataset* entity strings to a list of ICPSR study name strings and find their closest matches. The distribution of match scores is normal and centered at zero, which we set as our threshold. If a predicted entity matched an ICPSR study name with a match score of zero or greater, we classified it as an ICPSR dataset reference. A match below this threshold was classified as a non-ICPSR dataset reference and matches that did not receive a score were morphologically distinct, suggesting that they were not dataset references. **Figure 7** shows the twenty-five most frequently occurring dataset entities matched to their most similar ICPSR study name.

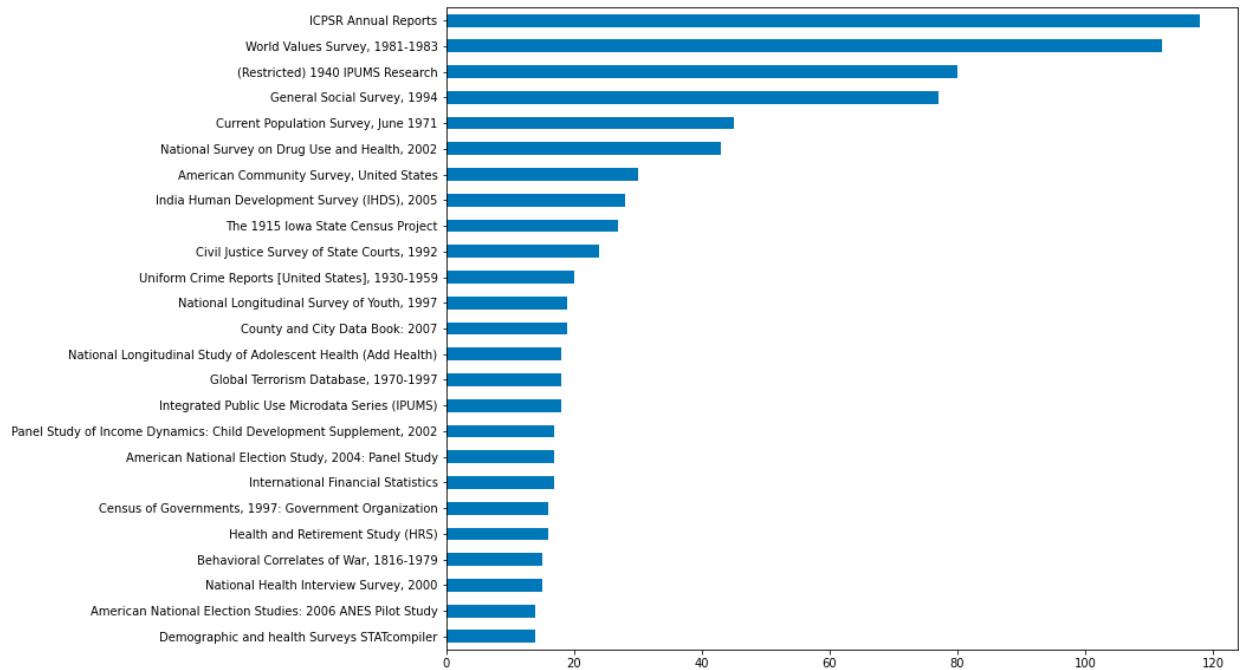

**Figure 7. Most frequently detected dataset entities matched to their closest ICPSR study names**

Using this method, we partitioned the extracted entities into three categories: ICPSR datasets (n=1,128 references found in 754 publications); non-ICPSR datasets (n=1,787 references found in 815 publications); and non-datasets (n=617 references found in 477 publications). References that matched ICPSR datasets were referred to the ICPSR Bibliography team for review, while those that did not match ICPSR study names were shared with the ICPSR Data Acquisitions team. Distinguishing between ICPSR and non-ICPSR data references can be used to notify ICPSR of potential data products that might be of interest for the archive to acquire in the near future. The model also does not discriminate between social science and other types of dataset references. We find that 67% (720/1,074) publications reference more than one type of dataset (ICPSR; non-ICPSR) and that 20% (691/3,535) of datasets are co-referenced with other datasets. Our pipeline supports co-citation analysis by extracting datasets that authors have referenced together in publications, regardless of their source domain.

To find where the predicted dataset references occur across publications, we normalized the section names (e.g., converted "Methodology" to "Methods") and counted the dataset references in each section. **Table 3** shows the sections of publications where dataset references matching ICPSR study names were identified most often. The majority of dataset references were found in the Introduction sections of publications; this contrasts with prior studies of data citation, which found more data references in Methods and Materials sections of articles (Boland et al., 2012; Mayo et al., 2016; Park et al., 2018). Data references likely serve different rhetorical and scientific purposes in different sections, and we include examples that illustrate these potential differences. For example, data references in Introduction sections of articles often contextualized and justified the selection of a dataset. References made in the Data and Methods sections described the specific data the authors used, the composition of the dataset, and procedures that the authors applied to the data. We also found examples where authors referenced data in their Discussion and Conclusions as supporting or contrasting evidence. Authors also described data quality and limitations, or describe plans to extend their analysis through future use of a dataset.

| Section | References | Examples |
|---|---|---|
| Introduction | 142 | "To fill this niche, the 'Mature Adults Cohort of the **Malawi Longitudinal Study of Families and Health (MLSFH-MAC)**' was initiated in 2012." |

| Section | References | Examples |
|---|---|---|
| Data | 89 | "As an extension to our analysis, we use another household survey, the **Russia Longitudinal Monitoring Survey of the Higher School of Economics (RLMS-HSE)**, a panel dataset consisting of various survey rounds from 1994 to 2016." |
| Conclusions | 76 | "Data from **ICPSR Study 6576** show that 21 percent of the counties in our data experienced a border change between 1870 and 1960." |
| Methods | 60 | "**The Detroit Area Study, 1995: Social Influences on Health: Stress, Racism, and Health Protective Resources** provides a sample of 1,139 respondents in the Detroit area." |
| Discussion | 58 | "The need for increased investment in drug abuse treatment programs was underscored in the **2008 National Survey on Drug Use and Health**, where 3.0% of the total population (i.e., 7.6 million persons aged 12 or older) were estimated to be in need of treatment for an illicit drug use problem." |

**Table 3. Sections of publications where ICPSR dataset entities were most prevalent**

In short, by extracting informal data references, we captured a wider array of scholarly work and extracted statements that reflect the diverse ways that researchers engage with datasets. Limited definitions of data reuse determine what counts as a data citation. Data reuse is often defined as use of data by someone other than the original creator of the dataset (Pasquetto et al., 2017; Zimmerman, 2008) in order to replicate an existing analysis (King, 1995) or to recombine existing data with new data. However, if a dataset is not listed in a references section or formally cited in the body of text, it is a matter of inference to determine the author's intention in mentioning a dataset. Strict collection requirements preserve only formal or explicit citations of research data; implicit or informal data references are lost. Future work should examine these differences to understand the specific intentions behind formal and informal dataset references.

**Limitations and outlook**
Even though machine learning can increase recall for data citations, gaps will remain because data citations do not capture all types of data reuse. Opaque data references persist, in which authors refer vaguely to datasets, variables, or instruments that they use in their research (Moss & Lyle, 2018). For example, datasets may be used, but are improperly cited, making references difficult to find. Researchers may also publish work in sources that are not formally recognized or reviewed, such as preprints, or mention research datasets in social media. As a consequence, an expanded ICPSR Bibliography will still not reflect the true level of engagement with research data. Capturing a more complete picture of engagement with research datasets as first class research outputs requires a broader view, which aligns with similar efforts to develop research altmetrics (Priem et al., 2012)).

In the present study, we have searched for literature referencing social science data in one scholarly database, Dimensions. We can expand search to include additional licensed databases and open science resources such as the xDD database (Peters et al., 2017) and the Project Freya PID graph (Fenner, 2019). This is motivated in part by licensing constraints we encountered when accessing full text in bulk. In addition, while we included reports and white papers in our training data, the majority of training data examples come from peer-reviewed journal articles. We would like to expand our training data to capture a wider variety of data references from theses, book chapters, and other materials that reflect a wider diversity of researcher career stages and disciplines.

While the NER model performs well by predicting sentences that contain the names of datasets, it cannot yet differentiate between instances of data analysis (e.g., using data to generate new insights or test new methods), making passing mentions to data products (e.g., providing background or describing a data product), and meta-level discussions (e.g., a critique of a dataset's limitations or quality). Cues from the text itself, such as the article section where the reference occurs, and the context, such as whether there are corresponding analytical tables and graphs, can help to distinguish types of data citations and their rhetorical functions. We plan to apply our pipeline to build a larger corpus of data statements in order to analyze the function and context of data citation. An extended analysis will shed light on the context and intent of data use.

## CONCLUSION

In this study, we have contributed: 1) a Named Entity Recognition (NER) model to detect informal data references; and 2) an expanded corpus of data-related social science literature, which can be used to analyze the context of research data reuse. We find that the NER model detects informal dataset references in academic text with relatively high recall. We also find that the pipeline increases ICPSR's capacity to automatically detect informal references made to their research datasets across a vast body of literature, complementing ongoing labor by staff to track research data use. We have expanded the ICPSR Bibliography of data-related literature, which supports long-range studies of data impact. Together, these contributions address gaps caused by inconsistent data citation practices and make it possible to detect and analyze data references at scale.

As data archives seek to maximize their impact across communities, data citations provide valuable evidence of the use and impact of research data. Our data detection pipeline also complements ongoing efforts to collect citations in two main ways. First, our approach reduces the amount of manual work performed by retrieving and reviewing publications. We also take advantage of human expertise to interactively train and update the NER model. The model also predicts and filters information, reducing the amount of content that staff review. The NER training pipeline uses human-in-the-loop feedback so that the time archive staff spend manually reviewing model predictions also provides high quality, expert training data to update the model and refine its predictions.

Second, by detecting informal data references at scale, we make it possible to gain a more comprehensive view of the data reuse landscape. We capture a wider array of mentions, references, and instances of data reuse. While the goal of the bibliography is to identify publications that use ICPSR data (e.g., for analysis, synthesis, and replication studies), our pipeline detects other types of discourse surrounding many kinds of data products. The analysis of data citations is an emerging area, given that data citation is a relatively new practice.

Detecting informal references to research data is needed to build robust data metrics. As more data is made available, an awareness of how users interact with data throughout its lifecycle can support the development of data-driven curation and collection policies. Understanding who is using data and for what purposes will also enrich data providers' and funders' understanding of the utility and impact of their research data. There is also potential to enhance and contextualize existing data metrics by providing more insights into the function of data citations and their intent. Data metrics that reflect a wider range of references to research data will better represent the diversity of data use and may also inspire novel reuse in the long-term.

## ACKNOWLEDGMENTS

Many thanks to David Bleckley and Elizabeth Moss of the MICA team at the Inter-university Consortium for Political and Social Research (ICPSR) for their support of this research. This material is based upon work supported by the National Science Foundation under grant 1930645.